# The role of Jupiter in driving Earth's orbital evolution: An update

Jonathan Horner[1,2], James B. Gilmore[2], and Dave Waltham[3]

[1] *Computational Engineering and Science Research Centre, University of Southern Queensland, West St, Toowoomba QLD 4350, Australia*
[2] *Australian Centre for Astrobiology, UNSW Australia, Sydney, NSW 2052, Australia*
[3] *Department of Earth Sciences, Royal Holloway, University of London*

**Summary:** In the coming decades, the discovery of the first truly Earth-like exoplanets is anticipated. The characterisation of those planets will play a vital role in determining which are chosen as targets for the search for life beyond the Solar system. One of the many variables that will be considered in that characterisation and selection process is the nature of the potential climatic variability of the exoEarths in question.

In our own Solar system, the Earth's long-term climate is driven by several factors – including the modifying influence of life on our atmosphere, and the temporal evolution of Solar luminosity. The gravitational influence of the other planets in our Solar system add an extra complication – driving the Milankovitch cycles that are thought to have caused the on-going series of glacial and interglacial periods that have dominated Earth's climate for the past few million years.

Here, we present the results of a large suite of dynamical simulations that investigate the influence of the giant planet Jupiter on the Earth's Milankovitch cycles. If Jupiter was located on a different orbit, we find that the long-term variability of Earth's orbit would be significantly different. Our results illustrate how small differences in the architecture of planetary systems can result in marked changes in the potential habitability of the planets therein, and are an important first step in developing a means to characterise the nature of climate variability on planets beyond our Solar system.

## Introduction

The question of whether we are alone in the universe is one that has long fascinated mankind. In the past twenty years, we have made the first steps toward being able to answer that question. In 1995, astronomers announced the discovery of 51 Pegasi b [1] – the first planet found orbiting a sun-like star. In the first years after that discovery, news of new exoplanets came infrequently, as planets very different to those in our own Solar system were discovered (e.g. [2][3][4]). As time has passed, the techniques by which we detect exoplanets have been refined (as detailed in [11]), and the temporal baseline over which observations have been carried out has expanded. As a result, astronomers are now beginning to discover planetary systems that more closely resemble our own – such as those hosting "Jupiter Analogues", giant planets moving on near-circular orbits that take of order a decade to complete (e.g. [5][6][7]). At the same time, improvements in the our ability to monitor the minute variations in a star's brightness that reveal the transit of an unseen planet across its host star have resulted in the detection of ever smaller planets. A thorough analysis of the data obtained by the *Kepler* spacecraft ([8][9]) has already resulted in the discovery of just over 1,000 planets[1], including several that are significantly smaller than the Earth (e.g. [10]).

It is now increasingly apparent that small planets are far more common than larger ones. This result is clearly seen in data taken by Kepler (e.g. [9][46][47]) and that obtained by radial-velocity search programs (e.g. [48][49]). Building on this work, a number of new exoplanet search programs will soon begin that should further bolster the number of small exoplanets known. These feature both ground-based programs (including dedicated facilities such as MINERVA [50] and NRES [51]) and space-based surveys (the Kepler K2 mission [52], TESS [53] and PLATO [54]).

With the rate at which new planets are being detected, and given the exciting new detection programs due to begin in coming years, it is likely that the first truly Earth-like planets (exoEarths) will soon be discovered. At that point, the search for life beyond the Solar system will become a key focus of astronomical research. However, the observations needed to characterise those planets, and search for any signs of life, will be incredibly challenging. The recent detection of water vapour in the atmosphere of a Neptune-sized exoplanet, HAT-P-11b [12] is a case in point. Despite the fact the planet is ~4 times the radius of Earth, and orbits its host star at around 1/20$^{th}$ the Earth-Sun distance, the detection still required the combination of 212 observations, totalling approximately 7 hours of integration time, using the Hubble Space Telescope's WFC3 G141 grism spectrometer. To gather spectroscopic evidence for life in the atmosphere of an exoEarth will be a significantly greater challenge, and would require more detailed and lengthy observations – making it a costly and time consuming process[2]. For that reason, it is imperative that astronomers select the most promising targets for the search for life before beginning their observations. But how will those targets be selected?

It is now thought that there are many factors that come together to render a given planet more or less habitable (e.g. [14], and references therein). One of the key factors to be considered is the stability of the climate of the planets in question. If all else were equal, it is reasonable to assume that the planet with the most stable climate would be the most promising target to search for life. It seems likely that otherwise ideal candidates could be rendered uninhabitable, should their climates vary too dramatically or rapidly for life to adjust and survive.

In our Solar system, the Earth's orbit is perturbed by the gravitational interaction with the other planets. This interaction results in Earth's eccentricity, inclination, and tilt varying over time. These orbital changes cause the radiation received at Earth's polar regions to vary over astronomical timescales, with periods of ~20 kyr and ~100 kyr. These cycles are known as Milankovitch cycles (e.g. [15][16][17]), and are thought to drive the glaciation/degalciation cycles observed over the last few million years (e.g. [18][19]).

---

[1] As of 18 Feb 2015, the number of confirmed planets discovered by the *Kepler* spacecraft stands at 1013 (http://kepler.nasa.gov), with a further 4,175 candidates awaiting confirmation.
[2] To illustrate the difficulty performing such observations, we note that recent simulations (as
[2] To illustrate the difficulty performing such observations, we note that recent simulations (as detailed in [13]), suggest that even the James Webb Space Telescope, due to be launched in 2018, will find it '*nearly impossible to measure spectra of terrestrial analogs*' (section 2.1.3).

Despite the role played by the Milankovitch cycles in driving Earth's climate and recent glaciations, the scale of the variations in Earth's orbital elements is actually relatively small. The same is not true elsewhere in the Solar system, however. For example, Mercury exhibits Milankovitch cycles with far greater amplitude than those of the Earth – with an orbital eccentricity (currently ~0.21) that can reach values as high as 0.45 (e.g. [20][21]). Were the Milankovitch cycles on the Earth equally extreme, it would sometimes move on an orbit that brought it closer to the Sun than Venus, at perihelion, and take it out to almost the orbit of Mars, at aphelion. The effects on Earth's climate might prove to be catastrophic, and it seems questionable whether life could survive, let alone thrive, on the surface of such a world.

In this paper, we present the preliminary results of a new study that aims to bring together *n*-body dynamical methods and climate modelling to assess the influence of planetary architecture on the climate of Earth-like planets. Here, we study the influence of the orbit of the planet Jupiter on the amplitude and frequency of Earth's orbital oscillations. Our work builds on that presented at the 13[th] Australian Space Science Conference [22] by using a greatly expanded suite of dynamical integrations (both in terms of number and duration), and using a corrected and improved methodology to perform the relativistic corrections required to model the orbit of the planet Mercury.

In the next section, we detail our methods, before presenting our preliminary results, and then conclude with a discussion of our plans for future work.

## Method

In order to study how the architecture of the Solar system influences Earth's orbital evolution, we use a modified version of the Hybrid integrator within the *n*-body dynamics package MERCURY ([23]). The standard version of the package models the dynamical interaction of test particles (both massive and massless) in a purely Newtonian sense. In addition, the modified version, developed for this work through the implementation of an additional user-defined force, takes account of the first-order post-Newtonian relativistic corrections [24]. This allows the code to accurately model the evolution of the orbit of the planet Mercury, when using a solely Newtonian method would yield erroneous results (e.g. [25][26]).

Using our modified version of MERCURY, we performed 159,201 simulations following the dynamical interaction of the eight planets in our Solar system for a period of 10 Myr. The initial orbits of all planets, except Jupiter, were held fixed throughout the suite of simulations, using the NASA DE431 [26] ephemeris to provide their current best-fit orbits. The simulation start epoch was taken as JD2450985.5, which corresponds to 00:00 21[st] June 1998, UT.

In each of the simulations, we placed the giant planet Jupiter on a different orbit. The inclination and rotation angles of Jupiter's orbit were identical in all cases, taken from the DE431 ephemeris. The semi-major axis and eccentricity of Jupiter's orbit, however, were varied from one run to the next. 399 unique values of semi-major axis were simulated, evenly spread across a 4 au region, centred on Jupiter's best-fit orbit (i.e. spanning the range 5.203102 ± 2.000000 au). At each of the semi-major axes tested, 399 discrete values of initial orbital eccentricity were considered. These ranged from a circular orbit (i.e. e = 0.0) to one with moderate eccentricity (e = 0.4), again in even steps. As such, the tested Jovian orbits spanned a 399 x 399 grid in *a-e* space, giving us our 159,021 simulations.

To maximise the accuracy of the orbital solution, a time-step of 1 day is used in each simulation. The orbital elements of each of the planets were output at 1,000-year intervals throughout the integrations. As a result of the large range of orbital elements considered for Jupiter, it was anticipated that at least some of the systems considered would prove dynamically unfeasible (as has often been observed when examining the dynamics of recently proposed exoplanetary systems – e.g. [27][28][29][30]). If any of the planets collided with one another, or with the central body, the simulation was stopped, and the time at which the collision occurred was recorded. Similarly, if any planet reached a barycentric distance of 40 au, that planet was considered ejected from the Solar system, and that integration was stopped, with the time recorded[3].

The results were used to create maps of the variability of the Earth's orbital elements as a function of Jupiter's initial semi-major axis and eccentricity. These maps, which build on earlier work studying the stability of proposed exoplanetary systems (e.g. [31][32]), give a quick visual guide to the degree to which the Earth's Milankovitch cycles are influenced by small changes in the orbit of Jupiter, and we present a number of examples of such plots in the next section. We are currently in the process of taking the numerical results of our simulations (the orbital elements for the Earth across the various runs) and using them as input for simple climate models (e.g. [33]), to examine how the observed variability in Earth's orbit might affect its climate. We anticipate that this analysis will be complete in the coming year.

## Preliminary Results

Figure 1 shows the variation in the Earth's orbital eccentricity (top panels, in red) and inclination (lower panels, in blue) for two exemplar simulations. The only difference between the two scenarios plotted was the initial semi-major axis chosen for Jupiter's orbit. In the left hand plots, Jupiter began the simulations at its true location in the Solar system, based on NASA's DE431 ephemeris ($a = 5.203102$ au). By contrast, the right hand plots show the scenario where Jupiter began the simulations at $a = 3.203102$ au. All other initial conditions were identical between these two simulations. It is immediately apparent that the amplitude of both the eccentricity and inclination excursions experienced by the Earth are broadly similar between the two runs. The maximum eccentricity in the scenario more closely resembling our own Solar system (left) was slightly higher than when Jupiter was closer to the Sun, but remained relatively small in both cases. However, the frequency of the cyclical behaviour in both eccentricity and inclination was much greater when Jupiter was placed closer to the Sun than when it began on its true orbit. Simply moving Jupiter inwards has had a significant effect on the Milankovitch cycles experienced by the Earth in these runs.

---

[3] A barycentric distance of 40 au was chosen for the 'ejection' distance as a reasonable compromise that allowed us to determine when the Solar system had destabilised. For any of the planets to reach 40 au would require a major re-arrangement of the system's architecture.

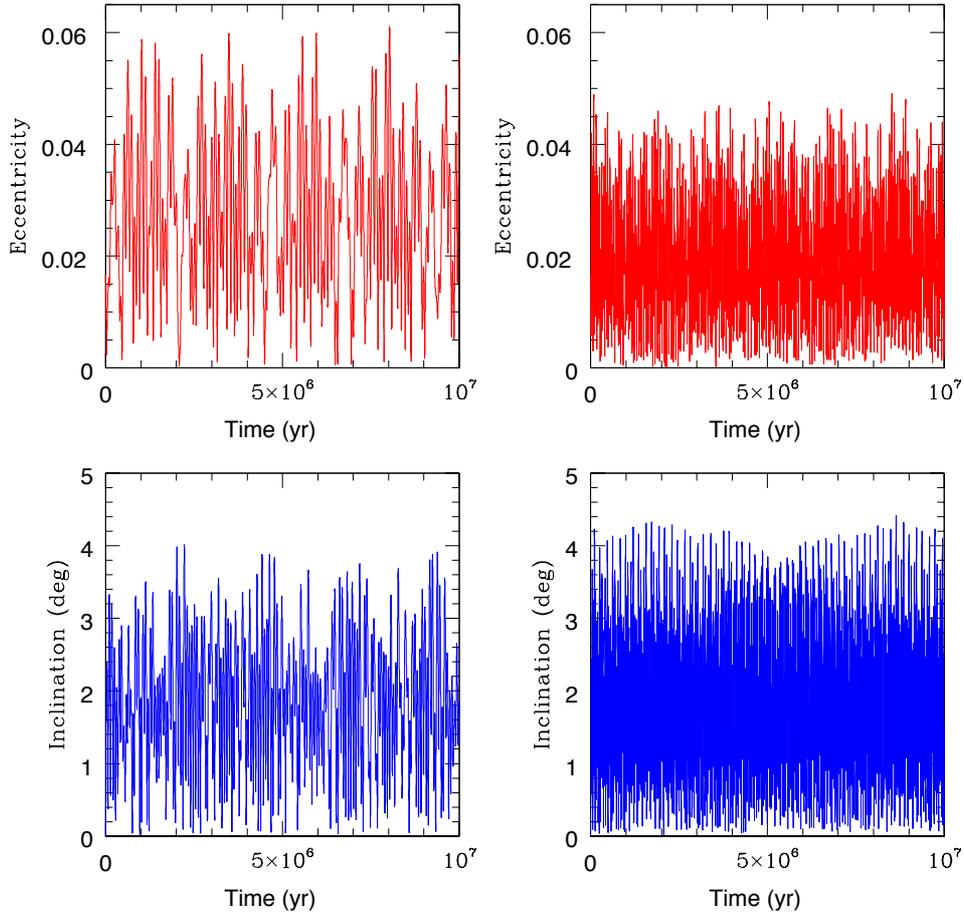

*Fig. 1: The variation in the Earth's orbital eccentricity (top, red) and inclination (bottom, blue), for two of the versions of our Solar system studied in this work. The left hand data is from the system that most closely resembled our own, whilst the right is for the scenario where Jupiter was shifted inwards by a distance of 2 au. All other initial conditions were identical between the two runs. It is clear that, though the amplitude of the variations in eccentricity and inclination were broadly the same between these two runs, the speed at which the variations occurred was greater for the scenario where Jupiter was closer to the Sun.*

The exemplar cases shown in Figure 1 clearly demonstrate how changing Jupiter's orbit can influence the Earth's Milankovitch cycles. However, changes to Jupiter's orbit can also have a much more dramatic effect – they can cause the Solar system to become unstable on very short timescales. As a result of the wide range of orbital architectures tested in this work, we found that a significant fraction of the Solar systems we created proved to be dynamically unstable, falling apart long before the end of our 10 Myr integrations. The stability of the Solar system as a function of Jupiter's initial semi-major axis and orbital eccentricity can be seen in Figure 2. Of the 159,021 versions of our Solar system we tested, almost 74% (117,549 systems) proved dynamically unstable within the ten million years of our simulations. The great majority of these featured Jupiters that were initially placed on orbits more eccentric than that displayed by our own Jupiter. However, there were values of initial Jovian semi-major axis where the Solar system was rendered unstable even for circular initial Jupiter orbits (such as those at around 4.25 and 4.95 au, where Jupiter and Saturn start in mutual 10:3 and 8:3 mean motion resonance, respectively). Similarly, there are two regions where the Solar system would remain stable even for moderately eccentric Jupiters – around 4.6 and 6 au. Once again, these are locations where Jupiter and Saturn would move on mutually resonant orbits – namely the 3:1 and 2:1 mean motion resonances, respectively.

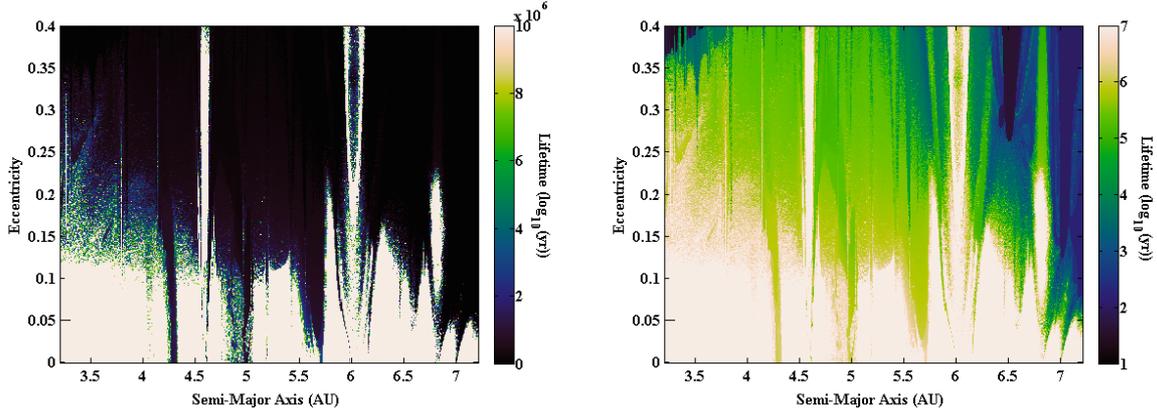

*Fig. 2: The stability of the Solar system as a function of the initial semi-major axis, a, and eccentricity, e, of Jupiter's orbit. In these integrations, the initial orbits of the other planets were held at their DE431 ephemeris values, and their evolution was followed for 10 Myr under the influence of their mutual gravitation. The left plot shows the lifetime of each system on a linear scale, whilst the right shows the same information on a logarithmic scale. It is clear that the majority of systems were unstable, even on the short timescales considered.*

The resonant features described above are clearly seen in the left panels of Figure 3, which shows the fraction of unstable simulations (top) and mean lifetime (bottom) of the Solar system as a function of Jupiter's initial semi-major axis and eccentricity. Both unstable and stable resonant features can be seen as troughs and peaks overlaid on a gradual trend to lower stability the closer Jupiter moves toward Saturn. The situation is more clear-cut when one considers the influence of eccentricity on the system's stability. Even scenarios featuring Jupiter on an initially circular orbit can prove unstable – but away from the destabilising influence of resonances (as seen in Fig. 2), the great majority of low-eccentricity solutions prove stable on the timescales considered in this work. Aside from those resonant regions, it is generally the case that scenarios where Jupiter has an initial orbital eccentricity less than ~0.125 prove dynamically stable – although this critical value is somewhat higher at low semi-major axes, and lower at greater semi-major axes. This transition region is evidenced in the right-hand panels of Figure 3 by the shoulder visible at $e \sim 0.1$, where the unstable fraction begins to climb more rapidly, and the mean lifetime begins to fall off more steeply.

For those regions where the system survived for the full 10 Myr of integration time, we have mapped the variability of the Earth's orbital elements as a function of time. In Figure 4, we show the variability of the Earth's orbital eccentricity over the integration period. In the left hand panel of that plot, we show the rms time variability of Earth's eccentricity, plotted on a logarithmic scale. In general, the closer to the Sun the initial orbit of Jupiter, the more rapidly the Earth's orbital eccentricity is driven to vary. As a second-order effect, the more eccentric the initial orbit of Jupiter, the more rapid are Earth's eccentricity excursions. This effect is particularly apparent in the strips of stability located at location of the 2:1 and 3:1 mean-motion resonances between Jupiter and Saturn.

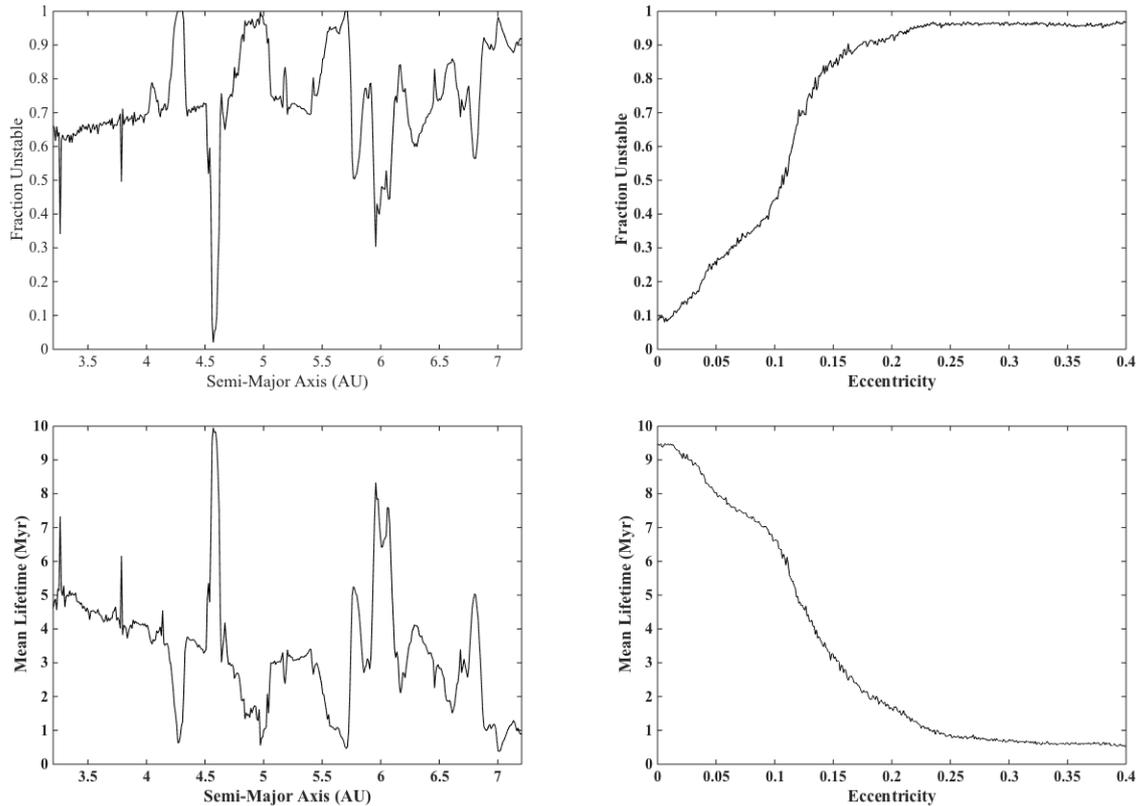

*Fig. 3: The fraction of our simulations that proved dynamically unstable (top) and the mean lifetime of those integrations (bottom), as a function of the initial semi-major axis (left) and orbital eccentricity (right) of Jupiter. The increasing instability of the systems tested as a function of orbital eccentricity can be clearly seen in the right hand panels, whilst the influence of mean-motion resonances (such as the 3:1 and 2:1 MMRs between Jupiter and Saturn, at ~4.6 and 6.0 au respectively) can be clearly seen in the left hand panels.*

The right-hand panel of Figure 4 shows the maximum eccentricity obtained by the Earth over the course of our integrations, again for those simulations where the Solar system survived intact for the full simulation time. In contrast to the rate at which Earth's eccentricity is driven to change, the scale of its maximum excursions seems unrelated to the initial orbital semi-major axis of Jupiter. Instead, it seems to be a function of the degree of stability of that planet's orbit. In general, towards the centre of the broad regions of stability, the maximum eccentricity obtained by the Earth's orbit is low, with it rising towards the edges of the stable regions – an indication that the Solar system is in the process of transitioning between a more stable and a less stable regime in those areas.

Figure 5 shows the variation of the Earth's orbital inclination over the course of our 10 Myr simulations. As with Figure 4, the left-hand panel shows the rms rate of change of Earth's orbital inclination, plotted on a logarithmic scale, whilst the right hand plot shows instead the maximum excursions seen in orbital inclination. As was the case with the rate at which Earth's orbital eccentricity was driven, it is clear from the left-hand panel of Figure 5 that the rate of inclination change is a strong function of Jupiter's initial semi-major axis. The closer that Jupiter orbits, the more rapidly the Earth's orbital inclination is driven. When taken in concert with the results shown in the left-hand panel of Figure 4, this is an indication that the frequency of the Milankovitch cycles might be a strong function of Jupiter's initial orbital location – the greater the Earth-Jupiter separation, the more slowly the cycles progress. Future work is necessary, however, before this conclusion can be confirmed.

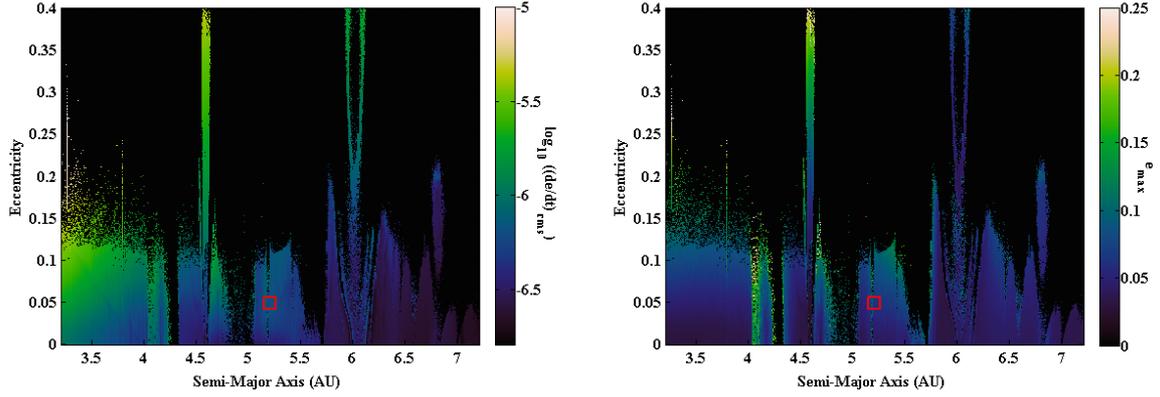

*Fig. 4: The variation of the Earth's orbital eccentricity, as a function of Jupiter's initial orbital semi-major axis and eccentricity. The left hand panel shows the variation in the root-mean-squared value of the rate of change of Earth's orbital eccentricity with time, whilst the right shows the maximum value that eccentricity obtained over the 10 Myr of our integrations. The red square in the plots shows the location of Jupiter in our own Solar system.*

The right hand panel of Figure 5 shows the maximum inclinations obtained by the Earth across the stable 10 Myr integrations we performed. The great majority of stable solutions show only very small excursions in inclination. There are, however, three bands (located at ~4.1, 4.7 and 6.3 au) where Earth's orbital inclination varies over a larger range. The band of values around ~15 degrees, at ~4.1, may well be the result of the 7:2 mean-motion resonance between Jupiter and Saturn, which would occur at ~4.14 AU. That narrow band at around 6.3 au is somewhat more mysterious, but given the apparent curved nature of the band (with an extension potentially visible at a ~6 au, e ~0.2) suggests that it may be the result of a secular resonance between the three planets in question (Jupiter, Saturn, and the Earth). Again, further work is needed to confirm or deny our suspicions in this case.

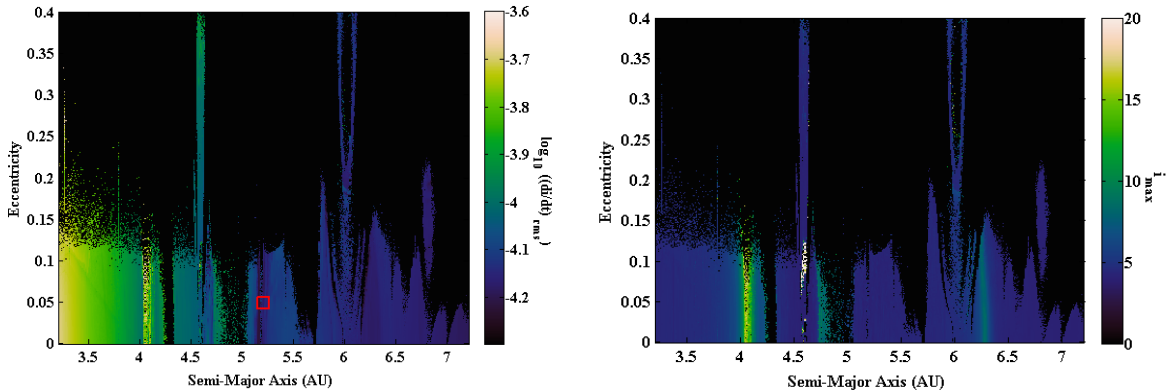

*Fig. 5: The variation of the Earth's orbital inclination, as a function of Jupiter's initial semi-major axis and eccentricity. Left: The RMS value of the rate of change of the Earth's orbital inclination with time. Right: The maximum inclination attained by the orbit of the Earth's over the 10 Myr of our integrations. The red square in the left-hand plot shows the location of Jupiter's orbit within our own Solar system.*

# Future Work

We are currently in the process of taking the orbital elements output for the many Earths in our integrations and using them as the input for detailed climate modelling, to determine how the changes in Earth's orbital variability (as illustrated in Fig. 2 and Fig. 4) would influence the climate of our planet. It is not necessarily the case that large excursions in orbital eccentricity and inclination would render the climate less clement for the development of life – clearly the rate at which variations occur will play an important role in determining the climatic response to the Milankovitch cycles themselves (e.g. [34]).

Once we have a firm handle on the interaction between the Solar system's architecture and the Earth's climate variability, we will look to extend our modelling to consider planets in the habitable zone of known exoplanetary systems. Our long term goal is to construct a systematic approach by which the potential habitability of such planets can be quickly assessed, so that the best possible candidates can be selected for the search for life elsewhere. Such studies will form a critical component of the search for life beyond the Solar system, and will complement studies of the other factors that can influence planetary habitability – ranging from stellar activity (e.g. [35][36][37]) to the impact regimes (e.g. [38][39][40][41]) and delivery of volatiles (e.g. [42][43]) that might be experienced by potential targets, and even the presence (or lack) of giant satellites (e.g. [44][45]).

Taken in concert, these studies will eventually provide a vital resource that will help observers to target those exoplanets that offer the greatest likelihood of hosting life like that seen on the Earth (e.g. [14]). Since studying the Milankovitch cycles will provide important information about potential exoEarth climate variability, we anticipate that astronomers will want to obtain as much information as possible prior to committing the extremely complicated observations that will be necessary to look for biomarkers on newly discovered planets.

# Acknowledgements

The work was supported by iVEC through the use of advanced computing resources located at the Murdoch University, in Western Australia.